\documentclass{aa}

\usepackage[utf8]{inputenc}
\usepackage{graphicx}
\usepackage{natbib}
\bibpunct{(}{)}{;}{a}{}{,} 

\usepackage{txfonts}

\usepackage{xcolor}

\newcommand{\corot} {CoRoT}
\newcommand{\kepler} {\emph{Kepler}}

\newcommand{\ds} {$\delta$ Scuti}

\newcommand{\muHz} {\mu\,{\rm Hz}}
\newcommand{\arma} {ARMA}

\newcommand{\conn} {{\cal C}_n}
\newcommand{\enn} {{\cal D}_n}
\newcommand{\disp} {\displaystyle}

\everymath{\displaystyle}
\newcommand{\eqn} [1] {
\begin{equation}
#1
\end{equation}}

\newcommand{\eqna} [1] {
\begin{eqnarray}
#1
\end{eqnarray}}

\begin{document}

\title{Inconsistencies in the application of harmonic analysis to pulsating stars}
\titlerunning{Inconsistencies in the application of harmonic analysis to pulsating stars}

   \author{J. Pascual-Granado
          \inst{1}
          \and
          R. Garrido
          \inst{1}
          \and
          J. C. Suárez
          \inst{2,1}
          }

   \institute{$^{1}$Instituto de Astrof\'{i}sica de Andaluc\'{i}a (CSIC), Granada 18008, Spain\\
   $^{2}$Dept. F\'{\i}sica Te\'orica y del Cosmos. Universidad de Granada. 18071 Granada. Spain\\
              \email{javier@iaa.es, garrido@iaa.es, jcsuarez@ugr.es}
             }

  \abstract
{Using ultra-precise data from space instrumentation we found that the underlying functions of stellar light curves from some AF pulsating stars are non-analytic, and consequently their Fourier expansion is not guaranteed. This result demonstrates that periodograms do not provide a mathematically consistent estimator of the frequency content for this kind of variable stars. More importantly, this constitutes the first counterexample against the current paradigm which considers that any physical process is described by a continuous (band-limited) function that is infinitely differentiable.}


   \maketitle
%

\section{Introduction} 

The unexpected huge number of frequencies found in multi periodic stars \citep{Poretti} and the ubiquitous presence of correlated noise in the residuals of the fitting of the light curves of pulsating stars, could have a common origin.        

The necessary condition for a correct Fourier analysis, i.e. to have a Fourier expansion, is guaranteed when the function is analytic, otherwise Fourier analysis is not a consistent approximation. In this work we examine the differentiability of the function describing  photometric data of pulsating stars obtained by \corot\ \citep{Auvergne} and \kepler\ \citep{Gilliland}. We determine how smoothly the data points of the light curve can be fitted in order to fully reproduce the function. We call this fine structure property the connectivity of the function at a given point. We studied this property by means of two numerical approaches: cubic splines - analytic\footnote{By definition only the first and second derivatives are continuous for cubic splines but we are interested here only in the first derivative. In any case, if the first derivative is not continuous the functions are non-analytic.}, and autoregressive moving average \citep{Box} methods - non-analytic.


\section{Analyticity}
Fourier analysis is a useful technique for frequency detection. In order to apply Fourier techniques to a given function, Parseval's theorem \citep{Parseval} must hold. This theorem states that the integral of the squared modulus of a function is equal to the integral of the squared modulus of its Fourier transform. For this to be true the function must be square integrable. The Fourier expansion of the function converges pointwise almost everywhere when it is square integrable \citep{Carleson}. Unlike the Fourier series, which is an infinite sum, the Discrete Fourier Transform (DFT) and its inverse, the IDFT, which is the analogous to the Fourier series, always converge if the sequence is finite, because they are finite sums. But when the Fourier series of the function is divergent, or not convergent to the value of the function at this point, the DFT no longer provides a mathematical description based on the frequency components of the function. That is, a frequency is only a well-defined physical variable when the Fourier series converges. Therefore, before a Fourier analysis can be applied to a time series, the convergence of the Fourier expansion of the function must be demonstrated.\par

On the other hand, the analyticity of a function is a sufficient condition for the square integrability \citep{analytic}. A function is said to be analytic if it is infinitely differentiable so that its Taylor expansion is convergent. If this condition is not met the convergence of the Fourier series is not guaranteed. That is, analyticity implies that Dirichlet conditions for convergence are fulfilled. 

In asteroseismogy studies (and often in astrophysics) such a convergence of the Fourier series is assumed, i.e. the analyticity of the underlying function is thus also assumed. To date, no methods to verify it have been proposed. Here we propose a practical method to verify the analyticity condition of the underlying function from the observed sampled data, which in consequence provide evidence of the convergence of the Fourier series.

In this sense, analyticity refers in this paper to the property that a given function can be expanded in Fourier series. To study the analyticity of a function we introduce a numerical approximation for the differentiability of discrete time series: the connectivity\footnote{Avoid confusion with graph theory term connectivity \citep{diestel}}.


\section{Connectivity}
Time-dependent astronomical data are given as a discrete sequence of samples from an underlying function, which is a representation of a given physical observable. This latter is understood as the physical quantity that satisfies all the properties\footnote{We want to stress here that an observable of a given physical property is not necessarily coincident with the function which represents it.}
 derived from the Lebesgue measure definition \citep{Lebesgue}. 
 
The discrete sampled sequence of photometric variations of luminosity and/or radial velocity measurements of pulsating stars is commonly called light curve (or RV).
Let us consider here such a series as a finitely close sequence of data points which samples a function composed by a set of infinitely close data points.
Under these conditions, we want to analyse whether such a light curve fully and unambiguously determines the properties of the sampled function. This is the well-known problem of sampling. 
It is known \citep[Nyquist-Shannon theorem,][]{Nyquist} that this is possible when the sampled function is band-limited to less than half the sampling rate,  i.e. the Nyquist frequency.
When this happens the complete information of the continuum (i.e. the sampled function) is contained in the discrete sequence. This implies that all the properties 
of the function can be studied, even those concerning the pointwise limits of the function. Consequently, we are allowed to adequately evaluate the differentiability of the sampled function. 

Let us now generalize our discrete sequence of data to sample a function composed by an stochastic sequence plus a deterministic contribution that can have or not a wave-like structure. Such decomposition is guaranteed under the hypothesis of stationarity by the Wold theorem \citep{Wold}. The random component might complicate the differentiability analysis of a time series, however the Kolmogorov continuity theorem allow us to fully characterize the properties of the function \citep{Revuz} in a similar way as it is done for pure deterministic signals. This requires necessarily the topological separability of the function,  which is applicable under certain technical constraints related with the moments of the data differences, particularly variations at most as those given by a power law (see Appendix A for a discussion on these constraints). When these restrictions are fulfilled Kolmogorov theorem states that a stochastic process has a continuous representation.

This is a necessary and sufficient condition for the connectivity analysis that we will define in the next section.

\subsection{Definition}
Given a data point from a discrete sequence, we can calculate a forward and backward extrapolation from the subsets bracketing the selected data point. Then, it is possible to check whether both extrapolations converge to the same value and coincide with the value of the selected data point.\par

The above theoretical framework permits the following self-consistent definition of the connectivity $\conn$ of a data point $y_n$ of a discrete sequence which samples the function $F(t)$
\eqna{C_n & = \epsilon^f_n - \epsilon^b_n\,,  \label{eq:defcn}}
where $\epsilon^f_n$, $\epsilon^b_n$, are the deviations of the forward and backward extrapolations from the sample $y_n$ 
\eqna{\epsilon^f_n & = y^f_n - y_n \nonumber \\
          \epsilon^b_n & = y^b_n - y_n\,.\label{eq:defeps}}
represented by $y^f_n$ and $y^b_n$, respectively. 
In fact, a discrete approximation of the derivability condition of F at the point $t_n$ can be:
\eqn{\frac{y^b_n-y_{n+1}}{\Delta t} = \frac{y^f_n-y_{n-1}}{\Delta t}\label{eq:pointwcond}}
where $\Delta t$ is the sampling rate of the sequence. Note that connectivity is closely related to derivability. In fact, we can define a new quantity $\enn$ as:
\eqn{\enn = \frac{\xi_n + y_{n+1}-y_{n-1}}{2 \Delta t}\label{eq:def_enn}}
which reduces to the typical point derivative \citep{derivative} for discrete data when $\xi_n$, the non-differentiable component of the sequence, is zero. Likewise, the connectivity as defined above is a function of $\xi_n$, that is:
\eqn{\conn = {\cal C}(\xi_n)+\epsilon_n}
From this equation, when connectivities are not zero but an independent normal stochastic sequence\footnote{This is the theoretically expected distribution of independent samples. It is usually understood as gaussian white noise.} they can be identified with the random term $\epsilon_n$ which is a numerical error. In this case the non-differentiable component is zero ($\xi_n=0$) and therefore the derivatives are still well defined. When connectivities are correlated with the signal the non-differentiable component is different from zero ($\xi_n\neq 0$) and therefore a derivative cannot be defined. In this case, the function is non-analytic and it does not satisfy the conditions under which Parseval's theorem is demonstrated, meaning that Fourier analysis is not consistent.

In summary, sampled data are obtained through the sampling of functions that needs not be necessarily differentiable a priori. Connectivities can be interpreted as jumps in the derivatives giving rise to discontinuities. This makes connectivities mathematically self-consistent.  They can be considered as equivalent to the non-differentiability coefficient introduced by \cite{Wiener} in a different physical context, and allowing to check the differentiability of the underlying functions.

\subsection{Method}
The forward and backward extrapolations necessary to obtain the connectivities have to be calculated numerically. Two conceptually different numerical approaches were considered. A comparison of both approaches allows us to confirm the results without ambiguity.

\subsubsection{Cubic splines}. The Stone-Weierstrass theorem \citep{Royden} states that if a function is uniformly continuous in a closed interval it can be approximated arbitrarily well, i.e. as closely as desired, by a polynomial of degree $n$,  $n$ being a natural number. Then, the entire function can be represented by a piecewise polynomial function of finite degree. But piecewise polynomial functions have a drawback, the Runge's phenomenon, showing oscillations at the edges of the approximated function. This is equivalent to the Gibbs phenomenon in the Fourier approximation. 

Taking all of this into account, we selected piecewise smooth polynomials for our modelling.  

Formulated in this general way any particular property can be extended to any analytic function. That is, if a continuous function can be fitted arbitrarily well with a given cubic spline parametric model then the analyticity condition is automatically satisfied, otherwise the convergence of the Fourier series is not guaranteed for that particular function. 
In short, the cubic splines approach gives a key information about local features of the function that generates the observed discrete sequence.

To calculate the coefficients of the cubic spline functions a tridiagonal linear system is solved~\citep{splines}. The only parameter affecting the spline interpolation that can be adjusted is the number of datapoints of the fitted segments. 

\subsubsection{ARMA} 
These models \citep{Box} give a parametric representation of a time series as a non-closed recursive formula. Autoregressive modelling has been used before for time series analysis in asteroseismology \citep{kovacs} although it has not become popular due to the difficulties for the physical interpretation of the fits, in contrast to the well-known Fourier frequencies.

We are interested here in the presence of any non-analytic content of the underlying function describing the observed signals. In contrast to the cubic splines, ARMA models are able to represent non-analytic functions, thereby allowing us to identify non-analytic signals from the residuals. 

The algorithm used to calculate ARMA models is similar to gap-filling algorithm MIARMA \citep{JPG}. This is based on minimizing rms residuals and hence, is insensitive to the arrow of time \citep{Scargle90}. However, here we extrapolate only one point each time and the segments of the modelled data are short enough to be able to estimate derivatives without bias. The invariance to the arrow of time allowed us to calculate backward extrapolations simply as an inverted forward extrapolation.

Like cubic splines modelling, the accuracy of the extrapolations with ARMA models depends on the number of data points of the segment to be modelled. In addition, ARMA extrapolations depend on two parameters: the orders $p$ and $q$, which are the number of autoregressive coefficients and the number of moving average coefficients, respectively. Hereafter, ARMA descriptions of the time series will be denoted by its coefficients as ${\rm ARMA}(p,q)$.
In general, an ARMA model is able to reproduce a sinus wave with only two terms. Accordingly, the number of terms increases with the frequency content of the signal.
The ARMA modelling requires of an iterative process involving the following steps:
\begin{enumerate}
        \item Identification of the order of the starting model using Akaike criterion \citep{Akaike} and described in detail in \citet{JPG}. This affects principally the velocity of convergence but does not modify the final result. 
        \item Calculation of the parameters using a Steiglitz-McBride algorithm \citep{Steiglitz} given the orders for the AR and MA contributions. 
        \item Evaluation of the validity of the model based on the goodness of the fit in backward and forward extrapolations. The algorithm starts again with increased orders until a minimum in the residuals is reached.
\end{enumerate}


\section{Connectivity analysis applied to stellar light curves}
The test was applied to the light curves of two delta Scuti stars observed by \corot\ and \kepler\ satellites: HD174936 and KIC006187665. The parameters required for this modelling are specific for each of the studied time series. However, in order to avoid misleading interpretation of spurious numerical effects we maintain the same parameters when characterising the connectivities for both time series. The \ds\ star HD 174936 was taken as a reference.

For the first step of the ARMA approach, the initial tentative orders were obtained using autoregressive methods (AR). Then we iterated until minimal randomly distributed residuals were obtained, fixing as optimal orders $p=20$ and $q=1$. Cubic splines and ARMA(20,1) connectivities were then calculated for every point in a segment of 1000 samples of the time series using subsets of 80 data points (40 forward and 40 backward extrapolations).  We adopted this number of samples per segment as a balance between computer time consuming and reliability of the ARMA modelling. In any case, we checked that the statistical properties of both approximations do not change significantly for a higher number of datapoints modelled.

In order to avoid artifacts we have selected data segments with no gaps and homogeneous sampling. Also, studied segments were statistically normalized allowing comparison of the connectivity results obtained using different time series.

\subsection{HD\,174936}
We first applied the method to the light curve of the multi periodic \ds\ star HD 174936 using as analytic solution the 422 independent frequencies found in its periodogram \citep{AGH}.

These are distributed in a range of frequencies below $900\,\muHz$. This very large range of excited frequencies is not predicted by any non-adiabatic model describing opacity-driven pulsators.

\begin{figure*}	
    \centering
    \resizebox{\hsize}{!}{\includegraphics{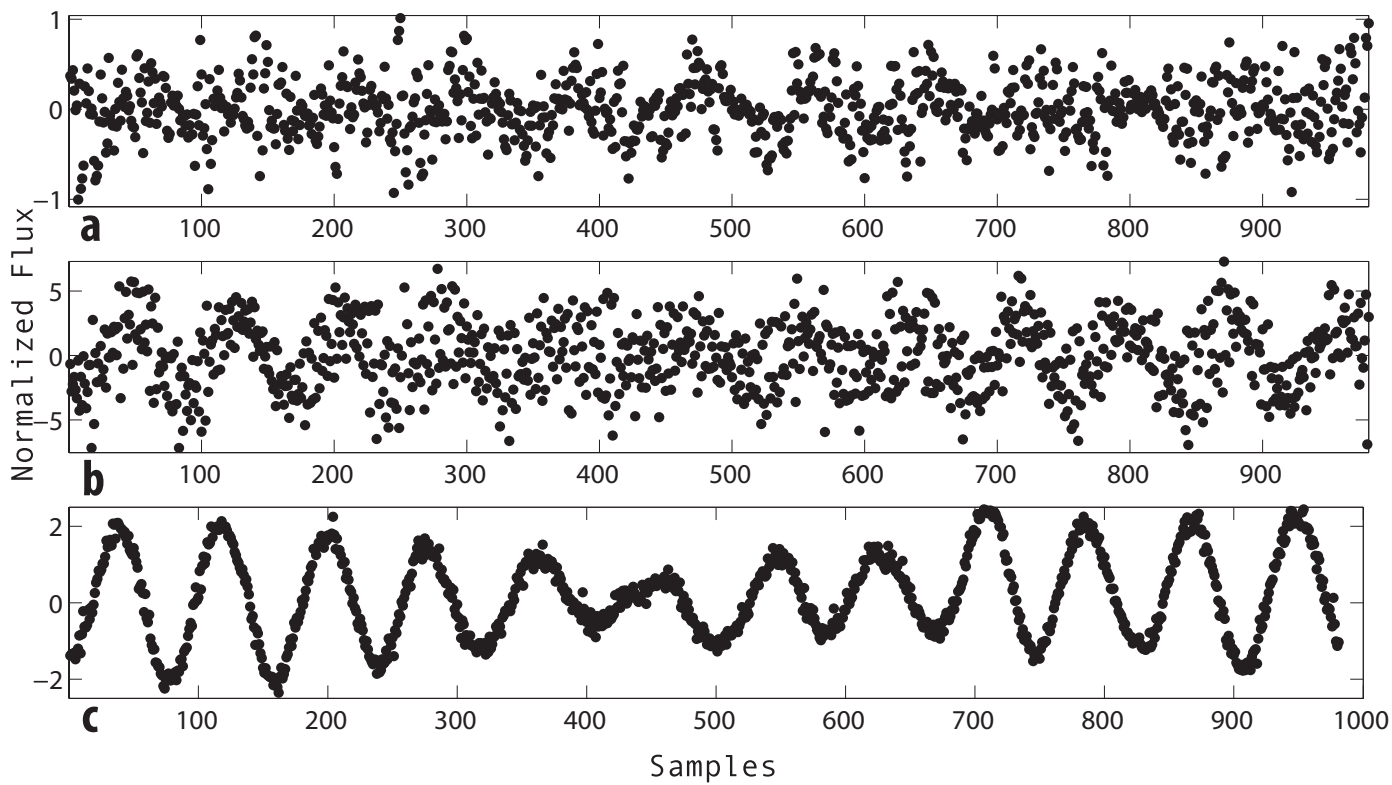}}
   \caption{Connectivities of the \corot\ data for the \ds\ star HD 174936. \textbf{a}, ARMA connectivities. \textbf{b}, cubic splines connectivities. \textbf{c}, the original light curve. (Note the different scaling factors in top and middle panels.)}    
   \label{fig:corot_conn1}
\end{figure*}
\begin{figure*}	
    \centering
    \resizebox{\hsize}{!}{\includegraphics{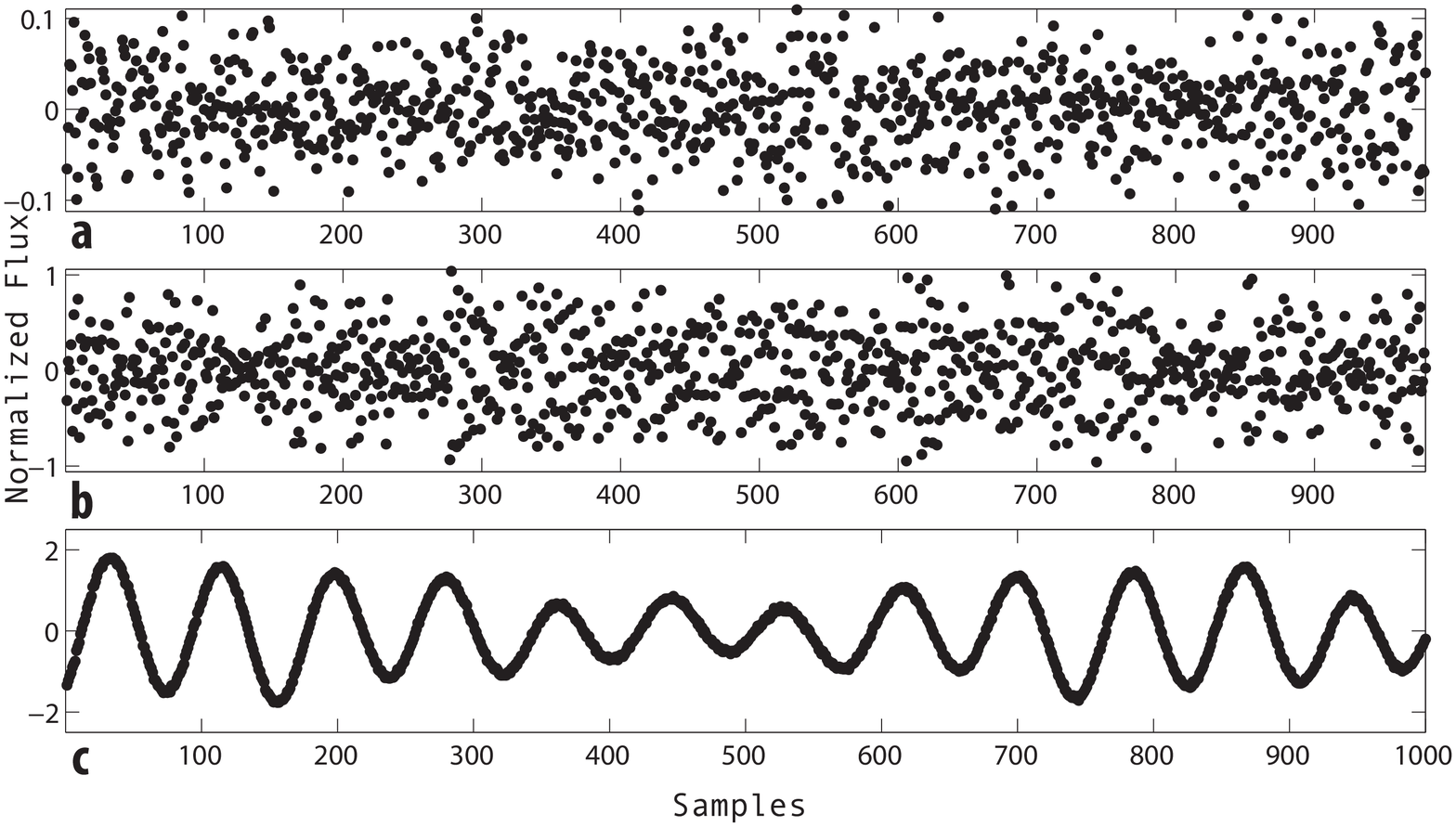}}
   \caption{Connectivities of an analytic model of the star HD 174936 based on the first 422 frequencies detected using standard Fourier techniques. \textbf{a}, ARMA connectivities. \textbf{b}, cubic splines connectivities. \textbf{c}, the original light curve. (Note the different scaling factors in top and middle panels.)}    
   \label{fig:corot_conn2}
\end{figure*}

The numerical methods described above were applied to an analytic model of this star. The time series of this model was calculated using the amplitudes and phases of the frequencies found in \cite{AGH} for the signal component, and a noise component based on additive gaussian white noise with a level obtained from the same reference.

For both the observed data (Fig.~1) and the time series of the analytic model (Fig.~2), the connectivities calculated using cubic splines have higher dispersion than the ARMA ones. Nevertheless, a striking effect is seen in the observed data that can not be noticed in the analytic model: spline connectivities are not randomly distributed but strongly correlated with the signal, suggesting that these data are not sufficiently well described by the analytic model. The same effect is present, although five times smaller in amplitude, for the ARMA results.

The difference between the analytic model and observed data has been usually interpreted as correlated noise caused by the effects of the turbulence present in the envelopes of stars \citep{Chaplin}. In order to check if the convection is at the origin of this phenomenon, some additional tests were performed by adding non-white noise to the analytic model following the procedure described in \citet{Kallinger}. A power law with two components (36 ppm and 19 ppm of amplitudes) was added in the frequency domain to the analytic model (see Appendix B).  The results of this test do not significantly change the previous ones, i.e. correlated connectivities were found. We conclude that the fully developed turbulence as represented by Harvey models \citep{Harvey} is not at the origin of this phenomenon.

\subsection{KIC\,006187665}
In order to discard possible instrumental effects specifically linked to a given instrument, a similar test was applied to the time series from another star KIC 006187665 observed by the \kepler\ satellite, and classified as a hybrid Gdor/Dscuti pulsator by \citep{Katrien}. 
The standard Fourier analysis of the light curve supplied by the satellite at a sampling rate of 60 seconds (short cadence regime),  yielded 659 significant peaks. 

The connectivities corresponding to the \kepler\ target show a very similar behaviour to the \corot\ ones (Fig.~3a,b), that is, cubic splines connectivities show a more dispersed distribution than the ARMA connectivities, and strongly correlated with the original \kepler\ time series. We have calculated the Pearson correlation coefficients for both splines and ARMA connectivities of the two time series studied here confirming these results (see table C1).

\begin{figure*}
\resizebox{\hsize}{!}{\includegraphics{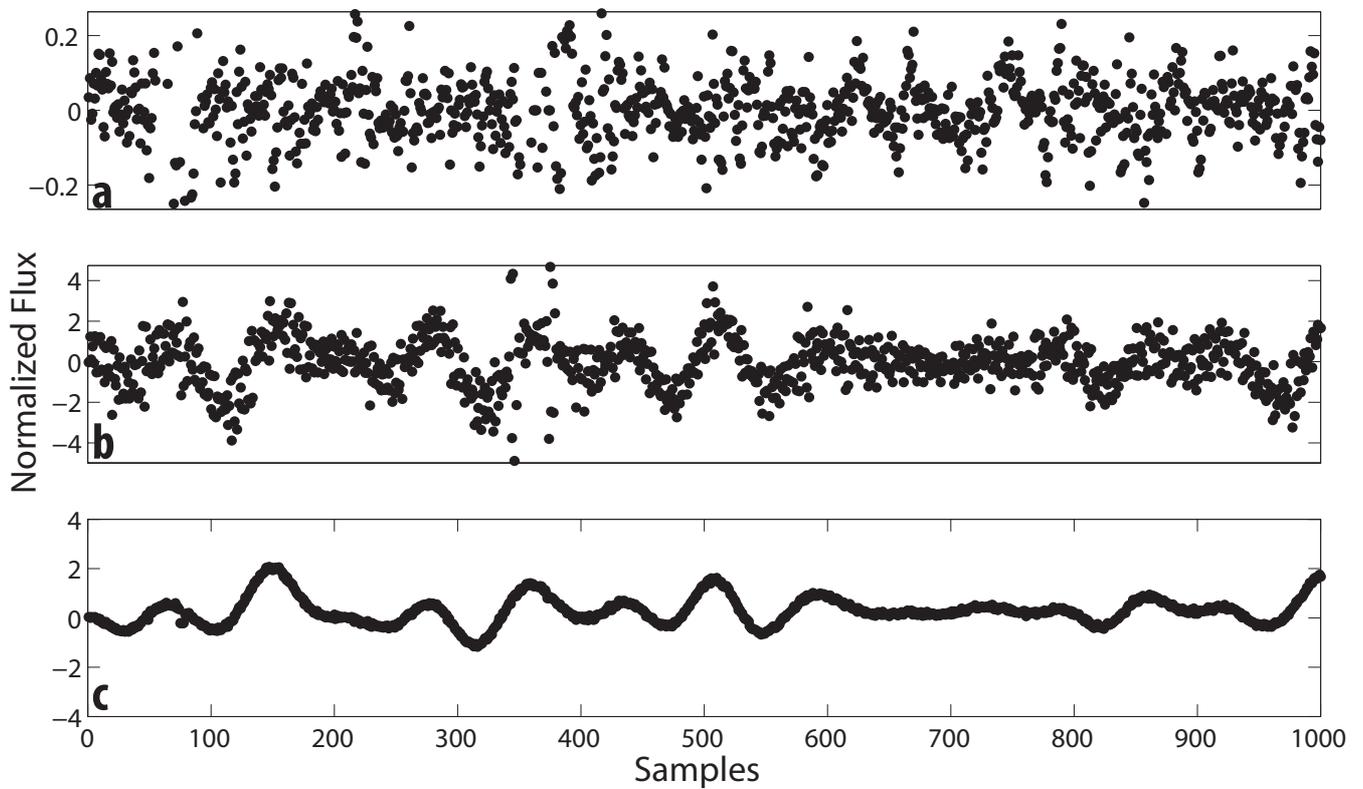}}
   \caption{Connectivities of the \kepler\ data for the hybrid star KIC 006187665 and the original light curve for comparison. \textbf{a}, ARMA connectivities. \textbf{b}, cubic splines connectivities. \textbf{c}, \kepler\ light curve for comparison. (Note the different scaling factors in top and middle panels.)}
    \label{fig:kepler_conn}
\end{figure*}

The non-random structures of the connectivities reflect the non-analytic component of the studied function. In order to characterize them, we calculated their amplitude spectrums (Fig.~C1) after verifying its applicability by using the test described in this paper (see Appendix D). 
The low frequency range in the case of splines connectivities is very similar to that found in the amplitude spectrum of the original time series. However a regular pattern can be seen along the frequency axis increasing in amplitude with frequencies.  The ARMA connectivities show some pattern at low frequencies whereas an almost flat spectrum, appearing as a white noise spectrum, is obtained at higher frequencies. The amplitudes of the ARMA connectivities are 1-2 orders of magnitudes lower than those calculated with cubic splines. 


\section{Discussion and conclusions}
The test introduced here through the connectivities allows us to fully characterise the differentiability of the underlying function of a time series. We remark that although the underlying function can be non-analytic due to the noise itself being non-analytic, our test assumes the non-analyticity of this component and evaluates the deterministic component through the correlation of the connectivities.

We have shown that at least for the stars studied here the functions underlying the observed time series are non-analytic. This comes from the finding that connectivities theoretically expected to be zero or a white noise distribution are strongly correlated with the signal. We have shown this to be an intrinsic property of those functions, which implies that their Fourier expansion convergence is not guaranteed. In addition, the fact that periodograms actually converge to a real value does not guarantee that this value is an asymptotically ($N\rightarrow\infty$) unbiased estimator of the spectral density of the underlying function. Therefore, the value estimated for the spectral density at a given frequency bin is not necessarily the true value.

In order to illustrate the possible relationship between the non-analyticity and the unexpected huge number of frequencies found in multi periodic stars, we performed a connectivity and frequency analysis of simulated data from a known function, the Weierstrass function \citep{Bailey}. This function is originally defined by an infinite sum of harmonic components (see Appendix E) but can be represented using eq.~\ref{eqn:weier2} with just a few components \citep{Weisstein}. However, it is considered a pathological function since it is continuous everywhere but differentiable nowhere. We show that when performing frequency analysis of the simulated data the residuals of the prewhitening process does not converge to white noise. Notice the large number of frequencies obtained as compared with the few components used to generate the function (see Fig.~\ref{fig:supp_weifreq}). This is an example of an inconsistency in the application of harmonic analysis that can be related to the non-analyticity of the underlying function.

We are not claiming that Weiertrass function is at the basis of the inconsistencies found in the time series of the two \ds\ stars  observed from space with different instrumentation, we are only using its pathological properties to try to "simulate" the real cases here studied. Although this peculiar function suggests a Fourier expansion in its standard form (eq.~\ref{eqn:weier1}) Fourier frequencies, given by an integer number of the argument, cannot be defined. This could points towards a reinterpretation of the term frequency in the context of the harmonic analysis of pulsating stars. The consequences of this new description of frequency are obviously out of the scope of this paper. Although the origin of these inconsistencies is unclear, our next step is to try to correct numerically the time series in order to overcome the analyticity test and therefore obtain a cleaner power spectrum.

We conclude that periodograms do not provide a mathematically consistent estimator of the frequency content for stellar variability of the objects here studied. Thereby, the concept of unambiguous detection of a physical frequency should be revised. This constitutes the first counterexample against the current paradigm which considers that any physical process is described by a continuous (band-limited) function that is smooth and infinitely differentiable.

This could be on the basis of the unknown nature of many phenomena related to the power spectrum of AF pulsating stars: the unexpected huge number of frequencies found in multi periodic stars and the range of these frequencies as mentioned before, and also the ubiquitous presence of correlated noise in the residuals of the fitting of the light curves.

This paper shows an inconsistency in the application of harmonic analysis to some pulsating stars observed by space missions like \corot\ or \kepler. This inconsistency is related with the non-analyticity of the underlying function, the origin of which is still to be unveiled.

\begin{acknowledgements}
The \corot\ space mission, launched on December 27th 2006, has been developed and is operated by CNES, with the contribution of Austria, Belgium, Brazil, ESA (RSSD and Science Programme), Germany and Spain. Funding for the \kepler\ Discovery mission is provided by NASAs Science Mission Directorate.
The authors acknowledge support from MINECO and FEDER funds through the Astronomy and Astrophysics National Plan under number AYA2012-39346-C02-01. J.P-G. acknowledges support from MINECO through the FPI grant number BES-2008-008252. JCS acknowledges support by the European project SpaceInn, with reference 312844 within the european SPACE program FP7-SPACE-2011-1, and also acknowledges funding support from the Spanish "Ministerio de Econom\'{\i}a y Competitividad" under "Ram\'{o}n y Cajal" subprogram.
\end{acknowledgements}

%
\appendix

\section[]{Kolmogorov continuity theorem}
As mentioned in Sect.3 this theorem \citep{Revuz} allow us to fully characterize the properties of a stochastic function through its continuous representation. The theorem comes after the following definition:\\\\
\textbf{\emph{Def}}.: An extension $(\tilde{Y}_t)$ of an stochastic process $(Y_t)$ is a process such that for every $t\geq0$, $P(Y_t=\tilde{Y}_t)=1$. 
As a corollary, if the process $(Y_t)$ satisfies for all times $T>0$  and $0\leq s,t \leq T$,
\begin{equation}
 \mathrm{E} \left( \| Y_t - Y_s \|^{\alpha} \right) \leq c \mid t-s \mid^{1+\varepsilon},  
\end{equation}
with $\alpha, \varepsilon, c$ positive constants, then there exists an extension of the process $(Y_t)$ that is a continuous process whose paths are almost surely continuous.
This is known as topological separability. 

The conditions above mentioned are quite mild and they allow us to consider stochastic processes as continuous paths instead of discrete values as it is the case for deterministic functions.

It can be demonstrated that these conditions are fulfilled by any white noise process and even for a non-gaussian noise having a spectral density proportional to $f^\alpha$ (with $\alpha<0$), i.e. the so-called coloured noise.

As an example, a real-valued Wiener process (so-called coloured noise proportional to $\nu^{-2}$) has continuous trajectories a.s.\footnote{almost surely in mathematical jargon} by this theorem. This means that it admits a  separable and progressively measurable extension \citep{Capasso}.

In any case,  in this work we refer to the stochastic component of a time series with the properties of gaussian white noise.

\section[]{The analytic model construction}
The numerical methods described in the main text were applied to an analytic model of the \ds\ HD 174936. This model was built using the amplitudes and phases of the first 422 frequencies found using a standard Fourier analysis in \citet{AGH}, fitted to an harmonic Fourier-like expansion of the form
\eqn{x^a (t_i) = \sum_{i=1}^n A_i\,cos(2\,\pi\,\nu_jt_i+\phi_j) + N^{\rm w} + N^{\rm nw}}
where $A_i$, $\nu_j$, and $\phi_j$ represent respectively the observed amplitudes,  frequencies, and phases. The $N^{\rm w} + N^{\rm c}$ terms represent the white noise, defined as normalised distribution scaled to the data,
\eqn{N^{\rm w} = < \sigma^{N}>}
and the non-white noise, calculated as 
\eqn{N^{\rm nw} = {\cal F}^{-1} [{\tilde N}^{\rm nw}]}
where ${\tilde N}^{\rm nw}$ is defined in the  transformed domain as
\eqn{{\tilde N}^{\rm nw}=\Bigg[\sum_{k=1}^m \frac{\disp a_k}{\disp 1+\bigg(\frac{\disp \nu}{\disp b_k}\bigg)^{ c_k}}\Bigg]^{1/2}e^{i\theta}}
where $a_k$ and $b_k$ are free parameters. The values of $c_k=4$ and $m=2$ were adopted following Kallinger \& Matthews (2010).
This noise was calculated setting the $\theta$ phases equal to a random uniform distribution in the [$-\pi,\pi$] interval. The light curve so constructed is supposed to be originated by the oscillation modes of the pulsating star plus granulation noise modelled as a fully developed turbulence regime \citep{Harvey} and given by eq.~B4.

\section[]{Pearson correlation coefficients}
\begin{figure*}
 \resizebox{\hsize}{!}{\includegraphics{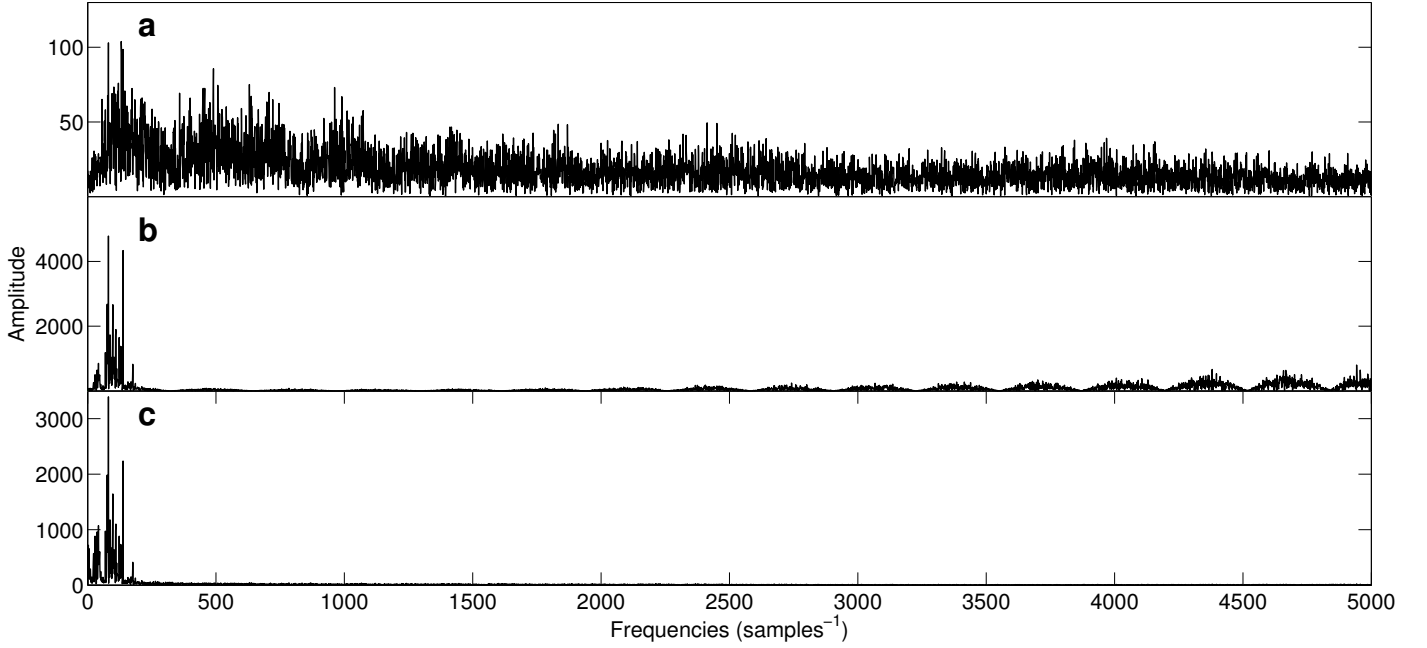}}
      \caption{Amplitude spectrum of the connectivities of the \kepler\ data for the hybrid star KIC 006187665. From top to bottom: panel (\textbf{a}) shows the amplitude spectrum of the ARMA connectivities, panel (\textbf{b}) amplitude spectrum of the cubic splines connectivities, panel (\textbf{c}) amplitude spectrum of the original light curve for comparison.}\label{spec}
\end{figure*}

 The correlations between the connectivities and their corresponding original time series are evident in Fig.\ref{spec}, even so, we calculated their corresponding standard Pearson coefficient in order to quantify the correlation between these two series (see Table \ref{tab:corr}).  We also calculated P-values for testing the null hypothesis  (i.e. no correlation). P-values were calculated using a t-Student distribution. Considering the $P_{\rm s}$ negligible values obtained for the time series of the selected stars,  the null hypothesis must be rejected under any computable limit of validity, i.e. the two series are strongly correlated. On the other hand, for the case of ARMA ($P_{\rm a}$), this limit can be put as an standard choice for this parameter, e.g. $P< 0.05$, i.e. only in 5\% of the cases the ARMA connectivities would be correlated with the original signal. 
 
Since ARMA models can capture the non-analytic component of the signal it is expected such non-correlation of the connectivities. 
 
\begin{table}
\caption{Pearson correlation coefficients ($\rho$) and the associated probabilities of no-correlation ($P$) for \arma\ and splines connectivities for the time series considered in this work. Subscripts a and s refer to the ARMA and splines approaches, respectively.}
  \begin{center}
  \vspace{1em}
   \renewcommand{\arraystretch}{1.2}
   \begin{tabular}[ht!]{|c|c|c|c|c|}
   \hline
    Time series & $\rho_{\rm a}$ & $P_{\rm a}$ & $\rho_s$ & $P_{\rm s}$ \\
     \hline
    CoRoT 7613 & -0.0075 & 0.8139 & 0.4243 & 5.8247e-45 \\
      \hline
    Analytic model & 0.0072 & 0.8199 & 8.371e-4 & 0.9789 \\
      \hline
    KIC 006187655 & -0.0257 & 0.4176 & 0.5553 & 5.8039e-82 \\
      \hline
   \end{tabular}
   \label{tab:corr}
 \end{center}
\end{table}

\section[]{Second-order connectivities}

Analyticity, understood as infinite differentiability, guarantees the convergence of the Fourier expansion of a function, so allowing to perform Fourier analysis. The test described in this work indicates whether the function underlying a time series is differentiable or not. Considering the connectivities as a new time series, the calculation of the connectivities of the connectivities (from now on second-order connectivities)  informs us about their differentiability.  If, using Pearson correlation coefficients, second-order connectivities are independent, normal,  and randomly-distributed,  then connectivities are differentiable. It is thus legitimate to perform a Fourier analysis of them. 

We applied this method presented  to the splines connectivities of the stellar light curve of KIC 006187665 (Fig.~D1, panel a). Contrary to the connectivities, second-order connectivities calculated using the splines approximation (Fig.~D1, panel b) show a normal white noise distribution. These, when calculated with splines show that the connectivities are differentiable. The ARMA approximation does not provide better fits than the cubic splines in this case.

\begin{figure*}
\resizebox{.9\hsize}{!}{\includegraphics{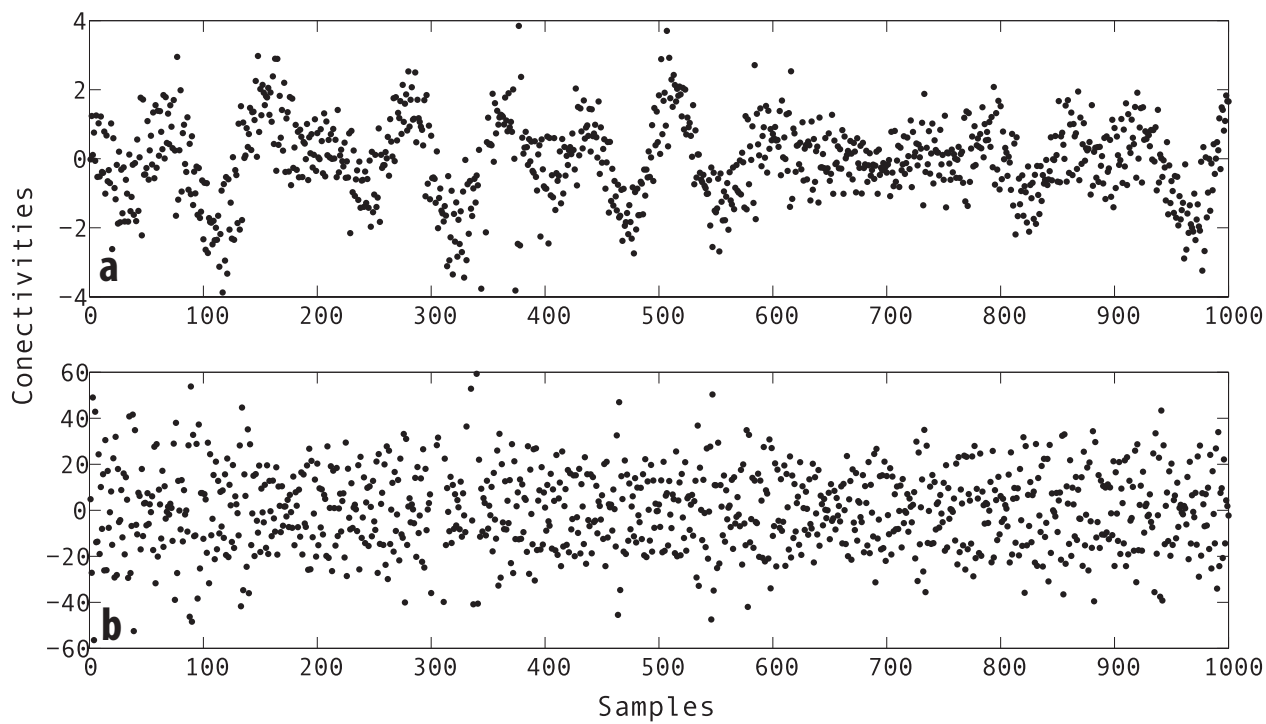}}
   \caption{Second order connectivities calculated using the cubic splines approximation for \kepler\ data of the the hybrid star KIC 006187665, and the first order connectivities for comparison. From top to bottom: Panel (\textbf{a}) first order spline connectivities, panel (\textbf{b}) second order splines connectivities. Note that the correlation present in the first is lost in the last.}\label{fig:figsupp_kic}
\end{figure*}

\section[]{Non-analytic model}
Frequency and connectivity analysis has been applied to a non-analytic model based on a numerical realization\footnote{obtained by sampling the continuous function} of the Weierstrass function. This is an extreme case of a function exhibiting non-analyticity because it is continuous everywhere but differentiable nowhere.

\subsection[]{Definition}
The original definition of the Weierstrass function is:
\eqn{W(x) = \sum_{n=0}^{\infty} a^n cos (b^n \pi x) \qquad 0<a<1, \qquad a\cdot b>1 + \frac{3}{2}\pi \label{eqn:weier1}}
and b, a positive odd integer number. For rational numbers $x=\frac{p}{q}$, which is our case,  $W(x)$ can be calculated with a finite sum such as:
\eqn{W \left(\frac{p}{q} \right) = \frac{\pi}{4 q^2} \sum_{n=1}^{q-1} \frac{\sin (\frac{n^2 p \pi}{q})}{\sin^2(\frac{n\pi}{2q})} \label{eqn:weier2}}
This formula allows to generate an exact numerical realization of the Weierstrass function without calculating an infinite sum. \par

For the non-analytic model that we used we generated three realizations of $W(x)$ with different \textit{q} values and obtained a 30000 datapoints time series by iterating on \textit{p} and adding each \textit{q} component.  

\begin{figure}
\resizebox{\hsize}{!}{\includegraphics{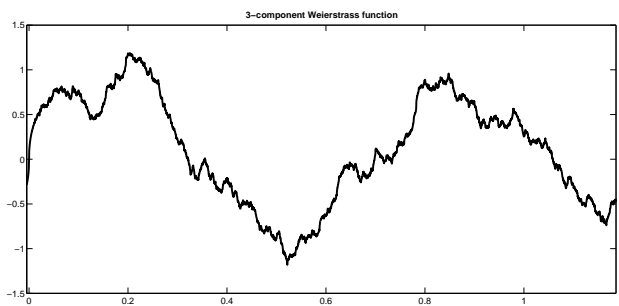}}
   \caption{A realization of the Weierstrass function with three components corresponding to three different q values as indicated in eq.\ref{eqn:weier2}.}\label{fig:supp_weidef}
\end{figure}

\subsection[]{Connectivity analysis}
\begin{figure*}
\resizebox{.9\hsize}{!}{\includegraphics{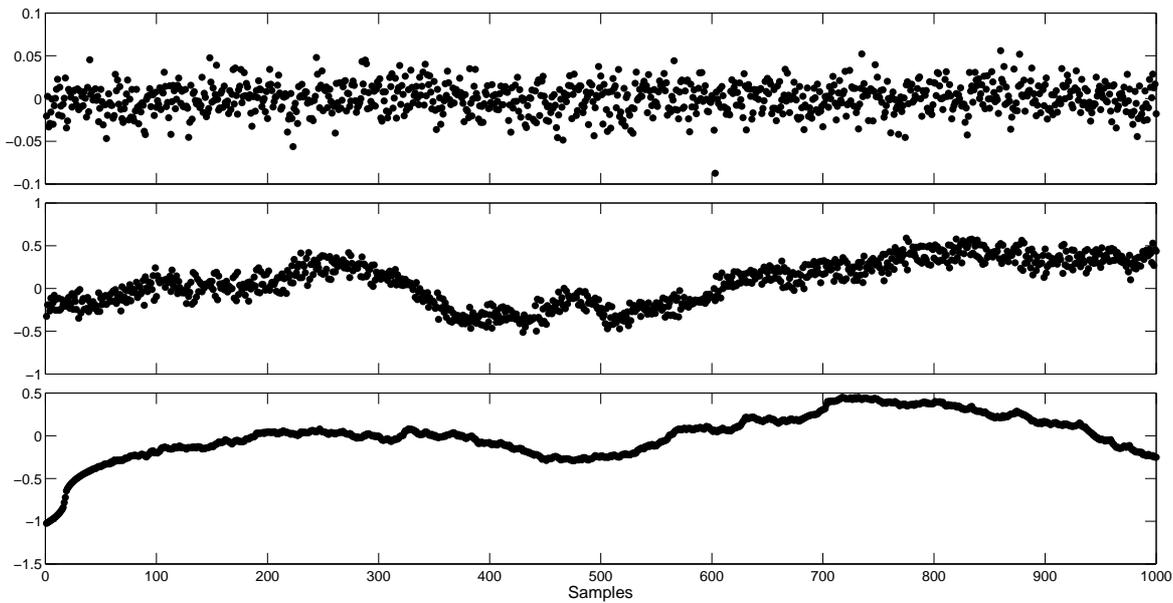}}
   \caption{Connectivities of a realization of a 3-component Weierstrass function. \textbf{Upper panel:} ARMA connectivities. \textbf{Middle panel:} cubic splines connectivities. \textbf{Lower panel:} the original light curve. (Note the different scaling factors.)}\label{fig:supp_weiconn}
\end{figure*}

We apply the numerical techniques introduced before to the time series generated as described in previous section. Connectivities calculated using the splines approximation are clearly correlated (Fig.\ref{fig:supp_weiconn}) and ARMA connectivities are not. These results are consistent with the definition and properties of the Weierstrass function and demonstrate with the analytic model discussed in  Appendix B the validity of the connectivity analysis as a test for the analyticity of the underlying function of a time series.

\subsection[]{Frequency analysis}
\begin{figure*}
\resizebox{.95\hsize}{!}{\includegraphics{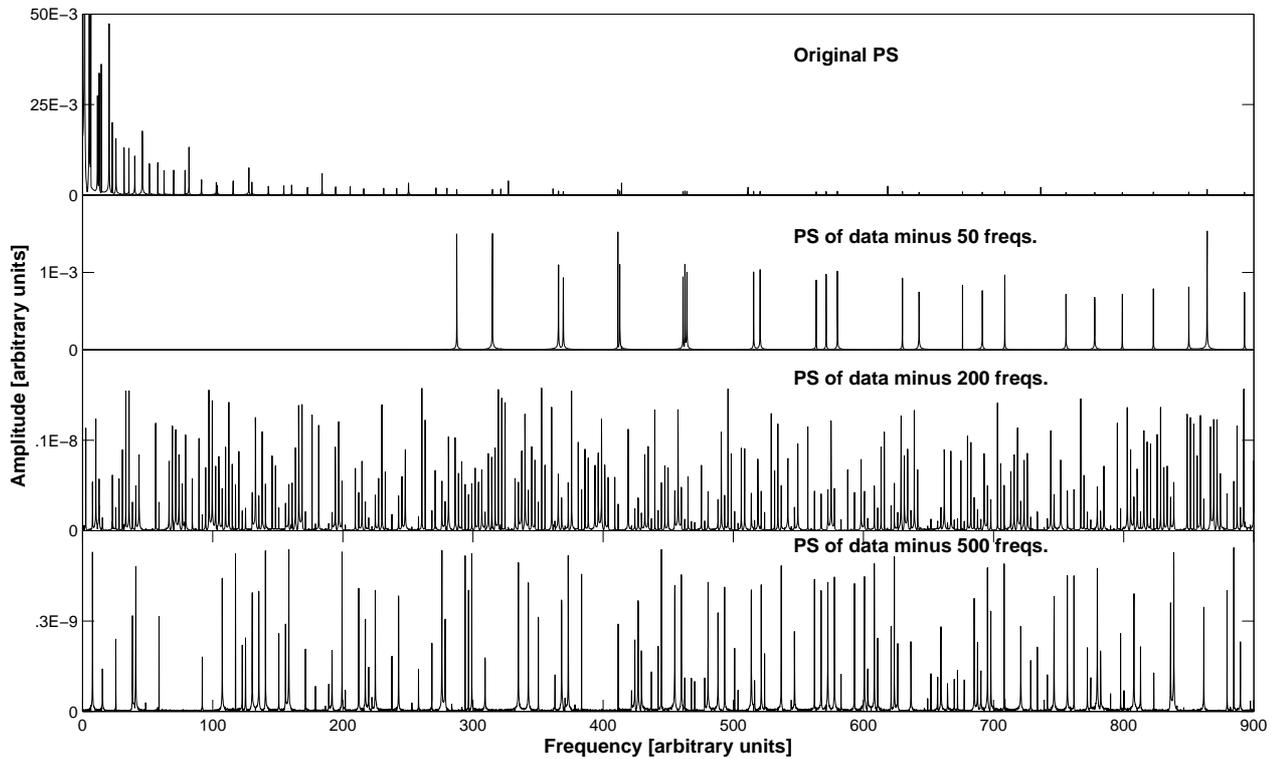}}
   \caption{Power spectrum of the time series generated with a realization of a 3-component Weierstrass function (top panel) and the power spectra obtained during different steps in the prewhitening process (succesive panels).}\label{fig:supp_weifreq}
\end{figure*}

A power spectrum of the time series of the non-analytic model and frequency detection was performed using \textsc{sigspec} \citep{sigspec}  in order to check the effect of the non-analyticity in a standard Fourier analysis. A sequence of power spectra was obtained during the prewhitening process of different sets of frequencies (Fig.\ref{fig:supp_weifreq}). The power spectrum at each step was significantly different to a white noise spectrum and the prewhitening process continued until the program stopped at 738 frequencies when the limit in significance was reached.\par
In conclusion, a frequency detection procedure using standard Fourier analysis is not convergent. This is due to the fact that, in this case, the power spectrum calculated with a DFT is not a consistent estimator for the original frequency content as a consequence of the non-analyticity of the underlying function. 

\bibliographystyle{aa}
\bibliography{connectivities}

\begin{thebibliography}{30}
\expandafter\ifx\csname natexlab\endcsname\relax\def\natexlab#1{#1}\fi

  \bibitem[Akaike(1974)]{Akaike} Akaike, H., 1974, IEEE Trans. Automatic Control, AC-19, 716-723

  \bibitem[Auvergne et al.(2009)]{Auvergne} Auvergne, M., et al. 2009, A\&A, 506, 411
  
  \bibitem[Bailey(2007)]{Bailey} Bailey, D. H., 2007, Experimental mathematics in action, Ak Peters Series
  
  \bibitem[Boor(1978)]{splines} Boor, Carl de, 1978, A Practical Guide to Splines, 1st ed., Series: Applied Mathematical Sciences, New York Berlin Heidelberg: Springer, 27
  
  \bibitem[Box(1976)]{Box} Box, George E. P., Jenkins, \& Gwilym 1976, M. Time Series Analysis: Forecasting and Control. San Francisco, CA. Holden-Day
  
  \bibitem[Burden(2011)]{derivative} Burden, R. L. \& Faires, J. D. 2011, Numerical Analysis, 9th ed. Boston, MA: Brooks/Cole
  
  \bibitem[Capasso(2005)]{Capasso} Capasso, V., Bakstein, D., 2005, An Introduction to Continuous-Time Stochastic Processes, Birkh\"auser      
  
  \bibitem[Carleson(1966)]{Carleson} Carleson, L. 1966, On convergence and growth of partial sums of Fourier series, Acta Math., 116(1), 135
  
  \bibitem[Chaplin(1997)]{Chaplin} Chaplin, W. J. et al. 1997, MNRAS, 287, 51-56
  
  \bibitem[Diestel(2005)]{diestel} Diestel, R., Graph Theory, Electronic ed. 2005, p.12
  
  \bibitem[Dijk (2009)]{analytic} Dijk, G. van  2009, Introduction to harmonic analysis and generalized Gelfand pairs, Berlin, New York. Walter De Gruyter
  
  \bibitem[Garc\'{\i}a Hern\'andez(2009)]{AGH} Garc\'{\i}a Hern\'andez, A. et al. 2009, A\&A, 506, 79
  
  \bibitem[Gilliland et al.(2010)]{Gilliland} Gilliland, R. L. et al., 2010, Publ. Astron. Soc. Pac., 122, 131
  
  \bibitem[Harvey(1985)]{Harvey} Harvey, J. 1985, Proc. ESA Workshop on Future Missions in Solar, Heliospheric \& Space Plasma Physics, Garmisch-Partenkirchen, 199
  
  \bibitem[Jech(1997)]{Lebesgue} Jech, T. J. 1997, Set Theory, 2nd ed. Berlin: Springer-Verlag
  
  \bibitem[Kallinger(2010)]{Kallinger} Kallinger, T. \& Matthews, J. M. 2010, ApJ, 711, L35-L39
  
  \bibitem[Kovacs(1983)]{kovacs} Kovacs, G. 1983, Solar Physics, 83, 123-128

  \bibitem[Parseval(1992)]{Parseval} Kaplan, W.  1992, Advanced Calculus, 4th ed. Reading, MA: Addison-Wesley 
  
  \bibitem[Pascual-Granado(2015)]{JPG} Pascual-Granado, J., Garrido, R., Su\'{a}rez, J. C., 2015, A\&A, 575, A78
  
  \bibitem[Poretti(2009)]{Poretti} Poretti, E., Michel, E., Garrido, R., et al. 2009, A\&A, 506, 85
  
  \bibitem[Revuz(1999)]{Revuz} Revuz, D. \&  Yor, M. 1999, Continuous Martingales and Brownian Motion, A Series of Comprehensive Studies in Mathematics (Springer),  293
  
  \bibitem[Reegen(2007)]{sigspec} Reegen, P., 2007, A\&A, 467, 1353

  \bibitem[Royden(1988)]{Royden} Royden, H. L. 1988, Real Analysis, Prentice-Hall
  
  \bibitem[Scargle(1990)]{Scargle90} Scargle, J. D. 1990, AJ, 359, 469 (1990).
  
  \bibitem[Shannon(1949)]{Nyquist} C. E. Shannon 1949, Communication in the presence of noise, Proc. IEEE, 37, 10
  
  \bibitem[Steiglitz(1965)]{Steiglitz} Steiglitz, K., \& L.E. McBride. 1965,  IEEE Trans. Automatic Control AC-10, 461
  
  \bibitem[Weisstein(2015)]{Weisstein} Weisstein, Eric W. "Weierstrass Function." From MathWorld--A Wolfram Web Resource. http://mathworld.wolfram.com/WeierstrassFunction.html
  
  \bibitem[Wiener(1923)]{Wiener} Wiener, N. 1923, Differential Space. J. Math. and Phys., 2, 131.

  \bibitem[Wold(1938)]{Wold} Wold, H.  1938, A Study in the Analysis of Stationary Time Series 2nd ed., Uppsala.  Almqvist and Wiksell

  \bibitem[Uytterhoeven(2011)]{Katrien} Uytterhoeven, K. et al. 2011, A\&A, 534, A125
 
\end{thebibliography}

\end{document}